\newif\ifsubmode
\submodefalse

\newif\ifprintfig
\printfigtrue

\ifsubmode
  \documentstyle[12pt,aasms4,epsf]{article}
  \received{}
  \accepted{}
  \journalid{}{}
  \articleid{}{}
\else
  \documentstyle[11pt,aaspp4,epsf]{article}
\fi

\lefthead{Gerssen et al.}
\righthead{Bar pattern speeds}

\newcommand{\etal}{{et al.~}}

\newcommand{\kms}{\>{\rm km}\,{\rm s}^{-1}}
\newcommand{\kmskpc}{\>{\rm km}\,{\rm s}^{-1}\,{\rm kpc}^{-1}}

\begin{document}

\title{Model-independent measurements of bar pattern speeds}

\author{Joris Gerssen}
\affil{Space Telescope Science Institute, 3700 San Martin Drive,
       Baltimore, MD 21218, USA}

\author{Konrad Kuijken}
\affil{Sterrewacht Leiden, PO Box 9513, 2300 RA Leiden, The Netherlands}

\author{Michael R. Merrifield}
\affil{School of Physics and Astronomy, University of Nottingham, 
   Nottingham NG7 2RD}



\ifsubmode\else
\clearpage\fi


\ifsubmode\else
\baselineskip=14pt
\fi


\begin{abstract}
The pattern speed in one of the fundamental parameters that determines
the structure of barred galaxies.  This quantity is usually derived
from indirect methods or by employing model assumptions.  The number
of bar pattern speeds derived using the model independent Tremaine \&
Weinberg technique is still very limited.  We present the results of
model independent measurements of the bar pattern speed in four
galaxies ranging in Hubble type from SB0 to SBbc.  Three of the four
galaxies in our sample are consistent with bars being fast rotators.
The lack of slow bars is consistent with previous observations and
suggests that barred galaxies do not have centrally concentrated dark
matter halos.  This contradicts simulations of cosmological structure
formation and observations of the central mass concentration in
nonbarred galaxies.
\end{abstract}


 
\clearpage


\section{Introduction}
\label{s:intro}

The rate at which a bar rotates, the bar pattern speed, is one of the
key parameters that controls the morphology and dynamical structure of
a barred galaxy.  Most determinations of the bar pattern speed rely on
indirect methods such as identifying morphological features with
resonance radii or by fitting models to observed velocity fields.  A
number of reviews describing these methods in detail can be found in
the proceedings of 'Disks of Galaxies: Kinematics, Dynamics and
Perturbations' (e.g. Teuben 2002).  A more direct, non-model
dependent method to estimate the bar pattern speed is due to Tremaine
\& Weinberg (1984, hereafter TW).  Kent (1987) and Kent \& Glaudell
(1989) were the first to apply this method to stellar
spectra. Merrifield \& Kuijken (1995) observed the same galaxy,
NGC~936, but used CCD spectroscopy to obtain spectra with
signal-to-noise ratios.  In subsequent years the TW method has been
applied to a few more barred galaxies: NGC~4596 (Gerssen \etal 1999),
NGC~1023 (Debattista \& Williams 2001) and NGC~7079 (Debattista \etal 2002).
Recently, Aguerri \etal (2003) presented an application of the the TW
method to a sample of five SB0 galaxies.

Interestingly, the results obtained with the TW method suggest that
the central regions in barred galaxies are not dominated by dark
matter halos, unlike their unbarred counterparts.  However, this could
be a selection effect because the TW method is best suited to gas poor
early type barred galaxies where the effects of dust extinction are
minimised.  Indirect techniques on the other hand usually rely on the
presence of gas and are therefore restricted to later type barred
galaxies.  Neither the TW method nor the indirect techniques may
therefore be representative of the whole population of barred
galaxies.

The number of barred galaxies that have been observed to date using
the TW method, mainly SB0 galaxies, is too small to unequivocally
ascertain whether centrally concentrated dark matter halos are truly
absent in barred galaxies.  However, even SB0 galaxies are not
completely devoid of dust.  In the absence of any quantitative
predictions of the effect of dust on TW measurements it may be
unnecessarily restrictive to apply this method to SBO galaxies only.
In this paper we present an analysis of the bar pattern speed in four
barred galaxies using the TW method.  Three galaxies, NGC~271,
NGC~1358 and ESO~281-31 are early type barred galaxies.  The fourth
galaxy, NGC~3992, an SBbc type is of somewhat later type and was
included to empirically assess how well the TW method can recover the
pattern speed in such a system.

\section{Tremaine \& Weinberg method}

Starting from the continuity equation TW showed that it is possible to
derive an expression that directly relates the pattern speed
$\Omega_p$ to two observationally accessible quantities: the luminosity
weighted mean velocities and the luminosity weighted mean densities
along lines parallel to disk major axis.  The underlying assumption is
therefore that the intensity is proportional to density.  In reality
the continuity equation is never strictly satisfied because of the
continuous formation of stars. However, in regular, early type spiral
galaxies this formation process is sufficiently slow that the TW
method can be applied.
	
In its original form the method is sensitive to centering errors and
suffers from low signal-to-noise ratios due to the --- relative to the
night sky --- low surface brightness of galaxies.  Merrifield \&
Kuijken (1995) used a revised implementation of the TW method that
circumvents these two drawbacks.  Briefly, this implementation
requires the determination of the {\it mean} line-of-sight velocities
and the {\it mean} positions of the stars along lines parallel to the
barred galaxy's major axis.  Merrifield \& Kuijken realised that a
luminosity-weighted mean velocity is naturally obtained by analysing
the stellar absorption lines of a longslit spectrum summed along its
spectral rows.  Moreover, this procedure also significantly increases
the signal-to-noise of the final spectrum.  The intensity profile of
the stars along a slit can be obtained by collapsing the spectrum
along the wavelength direction.  Both observables required by the TW
method can therefore be obtained from the same data set.  With this
implementation of the TW method the pattern speed follows from the
slope of a linear fit to the mean velocities and mean intensities of
longslit spectra obtained parallel to the major axis of the disk.
\begin{equation}
\Omega_p \sin i = \frac{\langle X \rangle - X_0}{\langle V \rangle - V_0}
\end{equation}
where $X_0$ is the luminosity center in the reference frame of the spectra
and $V_0$ is the systemic velocity of the galaxy.

The required observations are essentially measurements of the
deviation from symmetry.  Galaxies with a bar major axis position
angle in between about 20 to 70 degrees from the disk major and minor
axis are therefore suitable for an application of the TW method.
Although the TW signature will have its maximum strength when this
angle is close to 45 degrees.

\section{Observations and Reduction}

Longslit stellar absorption line spectra centered around the Mg b
triplet near 5200 \AA \ and with a spectral resolution of $\sim 5000$
were obtained for four barred galaxies.  Three galaxies, NGC~271,
NGC~1358 and ESO 281-31, were observed from La Silla on the nights of
August 28 and 29 2000 using the longslit spectroscopic mode of the ESO
Multi-Mode Instrument (EMMI) on the New Technology Telescope (NTT).  A
fourth galaxy, NGC~3992, was observed from La Palma on the night of
January 30 2000 using the blue arm of the ISIS spectrograph on the
William Herschel Telescope (WHT).  Both combinations of telescope and
spectrograph have rather similar characteristics in terms of overall
efficiency and dispersion per pixel.  Specifically, for the NTT data
the dispersion per pixel is 17.3 $\kms$ and the pixel scale is 0.54
arcsec, for the WHT data the numbers are 13.3 $\kms$ and 0.8 arcsec
respectively.  The data for all four galaxies are therefore of
comparable quality.

The galaxies and their relevant properties are listed in
table~\ref{t:sample}.  Position angles were derived from an isophotal
analysis of Digital Sky Survey images using the task ELLIPSE in IRAF.
Contour plots of the four galaxies with the positions of the slit
overlaid are shown in fig~\ref{f:mosaic}.  The slits were positioned
parallel to the disk major axis of each galaxy and centered on either
the nucleus or toward the bar edges.  In two cases (NGC~3992 and
ESO~281-31) we also obtained spectra along positions intermediate
between the center and the end of the bar.  A complete log of the
observations is given in table~\ref{t:obslog}.

All spectra were reduced in IRAF.  The exposures were bias
subtracted and flatfielded.  Small corrections for vignetting near the
slit-ends were also applied.  However, in practice the target galaxies
were all significantly smaller than the slit length.  The spectra were
wavelength calibrated and binned onto a logarithmic wavelength scale
using arclamp calibration frames taken before or after each science
exposure.  If necessary, contaminating light of foreground stars was
removed by linearly interpolating over these stars.  The sky
background was subtracted using the rows near the ends of the slit.
Finally, all rows in single longslit spectrum were summed to a
one-dimensional array excluding the outer most rows.  After sky
subtraction the rows toward the ends of the slit contain only
noise. Including these rows lowers the signal-to-noise of the summed
spectrum to a point were no meaningful velocity analysis can be
performed.

Stellar template spectra were also obtained on both observing runs.
Spectra of the K0III stars HD210396 and HD168424 were obtained on the
NTT.  On the WHT run spectra were taken of the template star HD107288
(K0).

\section{Velocity Analysis}
\label{s:velanal}

Stellar velocities were derived using the standard assumption that the
observed galaxy spectrum is the convolution of a typical stellar
template spectrum (usually a K giant) with the line-of-sight velocity
distributions (LOSVD).  Several techniques have been developed over
the years to obtain the LOSVDs from galactic spectra. The standard
cross correlation technique (Tonry \& Davis, 1979) is among the most
widely used such methods but assumes that LOSVD can be approximated by
a Gaussian.  We used an updated version of this method, the IRAF task
XCSAO (see Kurtz \& Mink 1998) to obtain estimates of the mean
velocities in the galaxies.  The continuum subtracted galactic spectra
where correlated against all continuum subtracted template spectra
that were available for each run.  The results, however, do not depend
significantly on the choice of template.  Before cross-correlating the
spectra, the data were Fourier filtered to suppress noise (large $k$
numbers) and residual continuum variations (low $k$ numbers).

An initial TW analysis of the spectra was also performed using XCSAO.
However, the distribution of stellar velocities in a disk galaxy often
deviates from a Gaussian distribution.  Especially in the tangential
direction, and the tangential velocity component dominates the
line-of-sight velocities along the major axis of a disk galaxy.

A more refined analysis was subsequently performed using the
Unresolved Gaussian Decomposition (UGD) method of Kuijken \&
Merrifield (1993).  UGD can measure deviations from Gaussian profiles
by assuming that the intrinsic line profile shape can be approximated
as a sum of Gaussians spaced at regular (user-defined) intervals.  The
only free parameters in the fit are the amplitudes of the individual
Gaussian components.  The advantages of a non-parametric method like
UGD are that it does not presuppose a particular form for the LOSVD,
and that it returns well-defined errors.  Following Merrifield \&
Kuijken (1995) we fit the LOSVD of each spectrum with three Gaussian
components uniformly placed at $V_{\rm sys} - 1.5\sigma_{\rm est}$,
$V_{\rm sys}$ and $V_{\rm sys} + 1.5 \sigma_{\rm est}$.  (Merrifield
\& Kuijken using data with higher signal-to-noise ratios were able to
fit two additional components at $-3\sigma$ and $3\sigma$
respectively.)  The systemic velocities $V_{\rm sys}$ as well as the
estimates of the velocity dispersions were obtained from the
cross-correlation analysis.  Although weakly, the derived LOSVDs do
indeed show the expected asymmetries (e.g. Merrifield \& Kuijken
1995).  The estimates of the pattern speed obtained with UGD are
therefore not significantly different from the XCSAO analysis.  For
data of higher signal-to-noise ratios these differences would be more
pronounced.  Besides a providing consistency check, the main advantage
of analysing the spectra with the UGD method is the robust estimate of
the errors.

\section{Results}

Along each slit the luminosity weighted mean velocity and the mean
light distributions were derived in the manner outlined above.  The
results are shown for all four galaxies in fig~\ref{f:results}.  The
errors in the intensity weighted mean positions along the slit are
negligible compared to the velocity errors and have therefore been
ignored in the analysis.  The derived velocities were not corrected to
heliocentric velocities since this is not required by the TW method.
Nevertheless the velocities are close to the published systemic
velocities with the exception of ESO 281-31. The literature value (see
table~\ref{t:sample}) for this galaxy differs by some $1500 \kms$ from
the value derived here. The published redshift is probably in error
since the NED database does not give a reference but simply states
that the redshift was derived prior to 1992.

The reliability of the velocity errors returned by the UGD algorithm
was assessed using a large set of artificially broadened, and noise
added template spectra .  The scatter in the mean velocities derived
from the broadened template spectra was found to be comparable to the
estimated velocity uncertainties.  The slope of the linear fits to the
mean velocities and the mean positions in each panel of
fig~\ref{f:results} yields $\Omega_p \sin i$ for each galaxy.  The
best-fit values and their $1\sigma$ errors are listed in
table~\ref{t:obspar}.

Multiple exposures per slit were obtained for all slits positioned
offset from the major axis.  These offsets were commanded to the
telescope by hand.  The spatial coincidence of multiple exposures that
were not obtained in sequence is therefore not perfect.  To test
whether this affects the derived pattern speeds we also derived the
mean velocities and mean intensities in each individual slit.  The
dashed lines in fig~\ref{f:results} show the linear fits to the
individual points (not shown to avoid crowding).  The derived slopes
are not significantly different from the values derived from the
combined spectra.

\subsection{NGC 271}

The data of this galaxy are consistent with a well-defined pattern
speed in this galaxy.

\subsection{NGC 1358}

Longslit spectra are available for the nucleus and on one side of the
bar.  Unfortunately, spectra on the opposite end of the bar could not
be obtained.  The pattern speed in this galaxy was derived using the
individual exposures only. The low number of data points did not
warrant an analysis on the combined spectra.  Consequently, the
derivation of the pattern speed is less secure in this galaxy than
it is for the other galaxies in the sample.

\subsection{ESO 281-31}

More longslit spectra have been obtained for this galaxy than for the
other galaxies observed with the NTT to facilitate a detailed
comparison with the stellar kinematics derived from a complete
Fabry-Perot data set in a forthcoming study.  The longslit data are
consistent with a well-defined pattern speed in this galaxy.

\subsection{NGC 3992}

The gas rich SBbc galaxy NGC~3992 has the latest Hubble type of all
galaxies in this sample. The star formation rate in this system will
therefore be higher than in the other three galaxies (e.g. Kennicutt
1998).  The basic assumption of the TW method is that the number of
tracers (i.e. stars) remains constant.  This condition is least well
satisfied in the case of NGC~3992.  Indeed, the scatter in the points
around the linear fit is larger than for any of the other --- earlier
type --- galaxies in this sample.  Several indirect studies of the
pattern speed in NGC~3992 have been attempted based on modeling the
gas kinematics.  Hunter \etal (1988) find a value for the pattern
speed in NGC~3992 of about 50 $\kmskpc$ and Kaufmann \& Contopoulos
(1996) derive a comparable value of 43.6 $\kmskpc$.  Using their
adopted distance of 14.2 Mpc we derive a pattern speed with the TW
method that is considerably larger at $83 \pm 5\kmskpc$.

\section{Co-rotation Radii and Bar Lengths}

A consistency check on the numerical values of the pattern speed can
be obtained by deriving the co-rotation radius in each galaxy,
i.e. the radius where the stars rotate with the same angular frequency
as the bar. Self-consistent bars cannot extent beyond their own
co-rotation radius and, consequently, the ratio, $\cal R$, of the
co-rotation radius to bar length cannot be smaller than 1.0.  The
circular velocities as a function of radius are required in order to
derive the co-rotation radius.  Unfortunately, published rotation
curves are only available for the gas-rich system NGC~3992.  Bottema
\& Verheijen (2002) have published an HI rotation curve for this system.
However, this rotation curve is only available for radii beyond the
bar.  We therefore extrapolated the rotation curve inward assuming a
constant value of $216 \kms$ (the average value obtained from their
table~3).  No rotation curves have been published for the other
galaxies in the sample. Global HI profiles are available for NGC~271
and NGC~1358 (Theureau \etal 1998) and can in principle be used to
estimate the circular velocities, assuming the rotation curve at large
radii is flat. 

The global HI profiles are rather noisy and we have therefore
attempted to measure the circular velocities directly from the stellar
absorption line spectra.  Although, these spectra were never intended
to derive spatially resolved kinematics, a cross-correlation analysis
(again using XCSAO) of the major axis spectra with the stellar
template spectra still yields meaningful estimates of the mean stellar
velocities as a function of radius, see fig~\ref{f:stellarkin}.  The
derived mean stellar velocities are constant at large radii.  However,
the amplitude of the flat part is somewhat lower than the circular
velocity.  This difference, the asymmetric drift, is accounted for by
the stellar velocity dispersion.  Our major axis spectra are of
sufficient quality to measure the central velocity dispersions,
$\sigma_0$, of each galaxy in our sample.  The central dispersions are
obtained from the Gaussian velocity distribution that, convolved with
a stellar template spectrum, gives the best match to the central row
of each longslit galaxy spectrum.  The radial behaviour of the
velocity dispersion cannot be ascertained from these spectra. Instead
we assume that the dispersions decline as $\sigma = \sigma_0
\exp(-R/2h)$, where $h$ is the photometrical scale length.  An
estimate of $h$ is obtained from the same photometry used to determine
the bar length (see section~\ref{s:barlength}) by measuring the slope
of the azimuthally averaged radial surface brightness profile outside
the bulge region.  The best-fit values of $\sigma_0$ and $h$ are
listed in table~\ref{t:obspar}.

With the above assumption for the radial behaviour of the velocity
dispersion and by assuming that the circular velocity is flat at large
radii the asymmetric drift can be calculated as follows (e.g. Gerssen
\etal 2000)
\begin{equation}
V^2_{\rm circ} - V^2_{\rm star} = \sigma_R^2 \left[ \frac{2R}{h} -
\frac{1}{2} \right],
\label{eq:vasym}
\end{equation}
where $\sigma_R$ is the radial component of the velocity dispersion.
However, from our observations we only have an estimate of the
line-of-sight velocity dispersions.  Rather than invoking additional
assumptions about the distribution of the velocity dispersion
components we identify $\sigma_R$ with the observed dispersions and
recognize that this approach yields an upper limit to the asymmetric
drift correction.  Using this approach we find that near the
co-rotation radii the asymmetric drift corrections are $\sim 10 \kms$
for both NGC~1358 and ESO~281-31 and $\sim 30 \kms$ for NGC~271.
Although not negligible, the upper limits to the asymmetric drift
corrections make only a small change to the derived co-rotation radii.
These radii and their uncertainties are listed in table~\ref{t:obspar}.

\subsection{Bar lengths}

\label{s:barlength}

The other piece of information required to derive the ratio $\cal R$
is the length of the bar.  To this end we obtained photometry for each
of the four galaxies in the sample from several sources.  $K$-band
photometry for NGC~271 was obtained from the Two Micron All Sky
Survey (2mass).  An $I$-band image of NGC~1358 was obtained from
from the Isaac Newton Group (ING) archive. $I$-band images of ESO
281-31 and NGC~3992 were kindly made available by Victor Debattista
and Roelof Bottema respectively.

Estimates of the bar semi-major axis length in each galaxy were
derived analogous to Aguerri \etal (2003).  With one exception the
techniques they use are based on Fourier decompositions of the
deprojected azimuthal surface brightness profiles.  The first estimate
of the bar length exploits the fact that along a bar the phase of the
$m=2$ component of a Fourier decomposition is constant (Debattista \&
Sellwood, 2000).  The bar length can thus be determined from a plot of
these phases as a function of radius.  The second method is based on
the ratio of the intensities in the bar and interbar regions (Aguerri
\etal 2000).  The bar and interbar intensities are defined as linear
combinations of the $m = 0, 2, 4$ and 6 terms of the Fourier
decomposition: $I_b = I_0 + I_2 + I_4 + I_6$ and $I_{ib} = I_0 - I_2 -
I_4 + I_6$.  Within the bar the ratio of the bar and interbar
intensities should be larger than $0.5 [\max(I_b/I_{ib}) -
\min(I_b/I_{ib})] + \min(I_b/I_{ib})$.  The left panels of
fig~\ref{f:barlength} show the results of the two methods applied to
the surface photometry of ESO 281-31.  With these data it also proved
possible to obtain the bar length from a bulge, disk and bar component
fit to the surface brightness profile along the bar major axis.  Both
the bulge and the disk were assumed to be exponential and the bar was
parametrised by the flat-type profile of Prieto \etal (2001).  The bar
lengths estimated with the three different method are similar to
within one arcsec (fig~\ref{f:barlength}).

The photometry available for the other three galaxies is of somewhat
lower quality and the bar length analysis in these galaxies was
therefore restricted to the Fourier decomposition methods only,
fig~\ref{f:barlengths}.  The adopted bar length for each galaxy listed
in table~\ref{t:obspar} is the average of the different methods used to
estimate this length.  In general, the different estimates agree quite
well and the associated errors are therefore small.  Only for NGC~3992
does the difference appear to be rather large.  The estimate based on
the ratio $I_b/I_{ib}$ is comparable to the bar length given by
Bottema \& Verheijen (2002).  However, the estimate based on the phase
of the $m = 2$ component yields a considerably smaller bar length in
this galaxy.  The surface brightness profile along the bar in NGC~3992
shows a clear break at a radius that coincides with the smaller of the
two estimates.  Perhaps the region between the break in the intensity
profile and the beginning of the spiral arm structure (which coincides
with the larger of the two estimates) is a region of gradual
transition from bar orbits to disk orbits.  Simulations show that such
regions are populated with chaotic orbits that can appear to either
support bar structure or spiral structure (Kaufmann \& Contopoulos
1996).

\subsection{Bar lengths 2: an alternative method}

The galaxies in our sample were all selected on the grounds of having
clearly identifiable bars in optical images.  A more quantitative
evaluation of the bar strength was made using the bar strength
analysis technique of Abraham \& Merrifield (2000).  This analysis
shows that the bars in our sample are all of comparable strength.
There is no hint of a correlation between the bar strength and the bar
pattern speed.  Kormendy (1979) suggested that bars slow down as they
become weaker.  The distribution of barred galaxies with measured
pattern speeds in Hubble Space (Merrifield 2002) is also not
inconsistent with this explanation.  However, the presently available
data does not support this interpretation, but the range in bar
strengths covered is still small.  Aguerri \etal (2003), using
literature classifications for the strength of the bars in their
sample, reach a similar conclusion.  Incidentally, the bar strength
analysis provides an additional estimate of the bar semi major axis
length (listed in the last column of table~\ref{t:obspar}).  All
derived bar lengths are consistent with the results presented in
section~\ref{s:barlength}.

\section{Discussion}

Once the the co-rotation radii and the bar lengths have been derived
it is trivial to determine the ratio $\cal R$.  The numerical values
for all four galaxies are given in table~\ref{t:twresults}.  Ratios of
$\cal R$ smaller than 1.0 are unphysical because they violate
self-consistency.  However, there is no a priori reason for the ratios
not to be significantly larger than 1.0.  N-body simulations of barred
galaxies (Debattista \& Sellwood 2000) as well as analytical arguments
(Weinberg 1985) suggest that the rate at which bars rotate is rapidly
(i.e. within a few orbital periods) slowed down due to the dynamical
friction exerted by the dark halo on the bar.  Bars with a ratio
${\cal R}$ of $\gtrsim 1.0$ are close to rotating as fast as they
physically can and are therefore said to be ``fast rotators''.  Bars
with larger values of ${\cal R}$ are called slow bars.  A somewhat
arbitrary, but now commonly employed practise in the literature is to
set the devision between fast and slow bars at a ${\cal R}$ ratio of
about 1.4 to 1.5.  Note that this does not imply anything about the
absolute value of the pattern speed of a bar.  To illustrate this
point the derived pattern speeds are converted to physical units
(column 3 of table~\ref{t:twresults}) by assuming the distances (using
an $H_0$ of $65 \kms$) and inclinations listed in
table~\ref{t:sample}.  These physical values show no correlation with
$\cal R$ but three out of the four galaxies have pattern speeds that
are consistent with fast rotators.  Although the errors on $\cal R$
are skewed toward large values, the maximum likelihood values of $\cal
R$ are clearly consistent with fast rotators for these three galaxies.
The fourth galaxy, ESO~281-31, is only just consistent with a fast
rotator. This system therefore warrants a more detailed study to
confirm whether the bar in this galaxy is a genuine slow rotator.

The results obtained from this sample of four galaxies agree well with
the previous determinations of bar pattern speeds using the TW
method. These studies all find that bars rotate fast.  The earlier
studies, however, were limited to the earliest type barred galaxies,
SB0 and SBa type galaxies.  A Compilation (Elmegreen \etal 1996) of
indirect results obtained for later type barred galaxies corroborates
the TW results. The average bar pattern speed obtained from this
compilation is $\langle {\cal R} \rangle = 1.2 \pm 0.2$.  A direct
conformation of the results relies on a successful application of the
TW method to later type barred galaxies.  The gas-rich system NGC~3992
was specifically included in the sample presented here to test the
feasibility of such an application.  The derived $\cal R$ ratio for
this system is formally consistent with a self-consistent bar.
Although the scatter in the derived velocities (fig.~\ref{f:results})
and the difference between the TW derived pattern speed and the
published --- indirect --- values suggests that this agreement is
perhaps fortuitous.  However, Aguerri
\etal (2003), find that two of the five SB0 galaxies in their sample
also have a $\cal R$ ratio smaller than 1.0 but argue that this just
reflects the scatter due to the uncertainties in the position angles
of the bar and of the disk.  Such uncertainties can lead to either
over- or underestimates of the derived pattern speed as simulations by
Debattista (2003) have shown.

An alternative explanation for the the small $\cal R$ value in
NGC~3992 is that it is due to dust absorption.  Intervening dust will
affect the derivation of the luminosity weighted mean velocities and
intensities. The derived pattern speed can therefore be quite
different from its intrinsic value.  This problem can be circumvented
observationally by using near IR observations since they are less
affected by dust obscuration.  The CO bandhead at $2.3\mu$ proofed to
be well suited to derive stellar kinematics and could therefore be
used for a TW type analysis in the near IR.  Baker \etal (2001)
attempted to do just that for NGC~1068 but found a ratio much smaller
than 1.0.  However, they only obtained two slit positions, which makes
it hard to constrain the pattern speed adequately.

The result obtained for the Sbc galaxy NGC~3992 is consistent with the
results obtained for SB0 systems (e.g. Aguerri et al 2003).  A
definitive test of whether the TW method is applicable to later type
--- gas rich --- barred galaxies therefore remains to be done. But
with the current availability of longslit near-IR spectrographs such a
test is now feasible.  The consensus reached from the studies of early
type barred galaxies, is that all bars appear to be fast rotators.
The dynamical friction between the bar and the dark matter halo must
therefore be small and this implies that barred galaxies do not have
centrally concentrated dark matter halos.  Navarro, Frenk \& White
(1997) show from cosmological simulations that galaxies should have a
universal dark matter halo distribution that is peaked in the center.
Indeed, (Courteau \& Rix 1999) find that nonbarred galaxies are
strongly dark matter dominated in their centers.  The results for
barred galaxies therefore appear to be at odds with their nonbarred
counterparts (e.g. Sellwood 2000).  Whether these results hold up for
later type barred galaxies depends largely on how well the TW method
can be applied to near-IR spectra.  An additional use of a large
sample of barred galaxies with measured kinematics is to test whether
these galaxies follow the Tully-Fisher relation (Courteau \etal
2003). This relation is based mostly on observations of nonbarred
galaxies, and can for these systems be explained in terms of the CDM
paradigm.


\acknowledgements

Based on observations collected with the NTT at the European Southern
Observatory of La Silla.  The WHT is operated on the island of La
Palma by the Isaac Newton Group in the Spanish Observatorio del Roque
de los Muchachos of the Instituto de Astrof\'isica de Canarias.  JG
thanks Roeland van der Marel, Victor Debattista and Roelof Bottema for
comments and discussions.  Much of the analysis in this paper was
performed using {\sc iraf}, which is distributed by NOAO.

\clearpage





\ifsubmode\else
\baselineskip=10pt
\fi


\clearpage


\ifsubmode\else
\baselineskip=14pt
\fi


\newcommand{\figcapmosaic}{
Contour plots of the four galaxies in the sample.  Foreground
stars present in the images were removed. North is to the top
and east is to the left in each panel.  The isocontours are chosen
such as to best delineate the orientation of the bar and the disk.  In
all four galaxies the position angle of the bar is intermediate
between the disk major and minor axis, as required by the TW method.
The solid lines indicate the positions along which we obtained the
stellar absorption line spectra.
\label{f:mosaic}}

\newcommand{\figcapresults}{
The derived mean line-of-sight velocities versus the luminosity
centroids for all slits in each of the four galaxies in the sample.
The slopes of the linear fits (solid lines) are a measure of $\Omega_p
\sin i$ in each galaxy. The dashed lines show the slopes derived from
individual exposures rather than from the combined spectra at each
slit location (see text).
\label{f:results}}

\newcommand{\figcapstellarkin}{
The stellar line-of-sight velocities derived using cross-correlation
for the three galaxies in our sample for which reliable estimates
of the rotation curves were not previously available.
\label{f:stellarkin}}

\newcommand{\figcapbarlength}{
Different estimates of the length of the bar in ESO~281-31.  The top
left panel shows an estimate based on the method of Debattista \&
Sellwood (2000): along the bar, the phase of the $m = 2$ component of
a Fourier decomposition of the (deprojected) surface brightness is
constant.  In the bottom left panel the bar length is estimated from
the ratio of bar/interbar intensities (see text and Aguerri \etal
2000). The dashed line indicates the location where this ratio is
consistent with the end of the bar (see text).  The panel on the right
shows a fit to the surface brightness profile along the bar major
axis. Both the bulge (short dash) and the disk (dotted) are assumed to
be exponential.  The bar component (long dash) is parametrised using
the flat-type profile of Prieto \etal (2001).
\label{f:barlength}}

\newcommand{\figcapbarlengths}{
Estimates of the bar semi-major axis length in the remaining three
galaxies of the sample.  The bar lengths are determined using the same
Fourier decompositions techniques as in Fig.~\ref{f:barlength}.
However, the photometry for these three galaxies is of lower quality
than the ESO~281-31 data, and reliable bulge/disk/bar component fits
to the surface brightness profiles could not be obtained.  In NGC~271
and NGC~1358 the images are over exposed in the centers. These regions
were therefore excluded from the analysis.
\label{f:barlengths}}


\ifsubmode
\figcaption{\figcapmosaic}
\figcaption{\figcapresults}
\figcaption{\figcapstellarkin}
\figcaption{\figcapbarlength}
\figcaption{\figcapbarlengths}
\clearpage
\else\printfigtrue\fi

\ifprintfig


\clearpage
\begin{figure}
\epsfxsize=15.0truecm
\epsfysize=15.0truecm
\centerline{\epsfbox{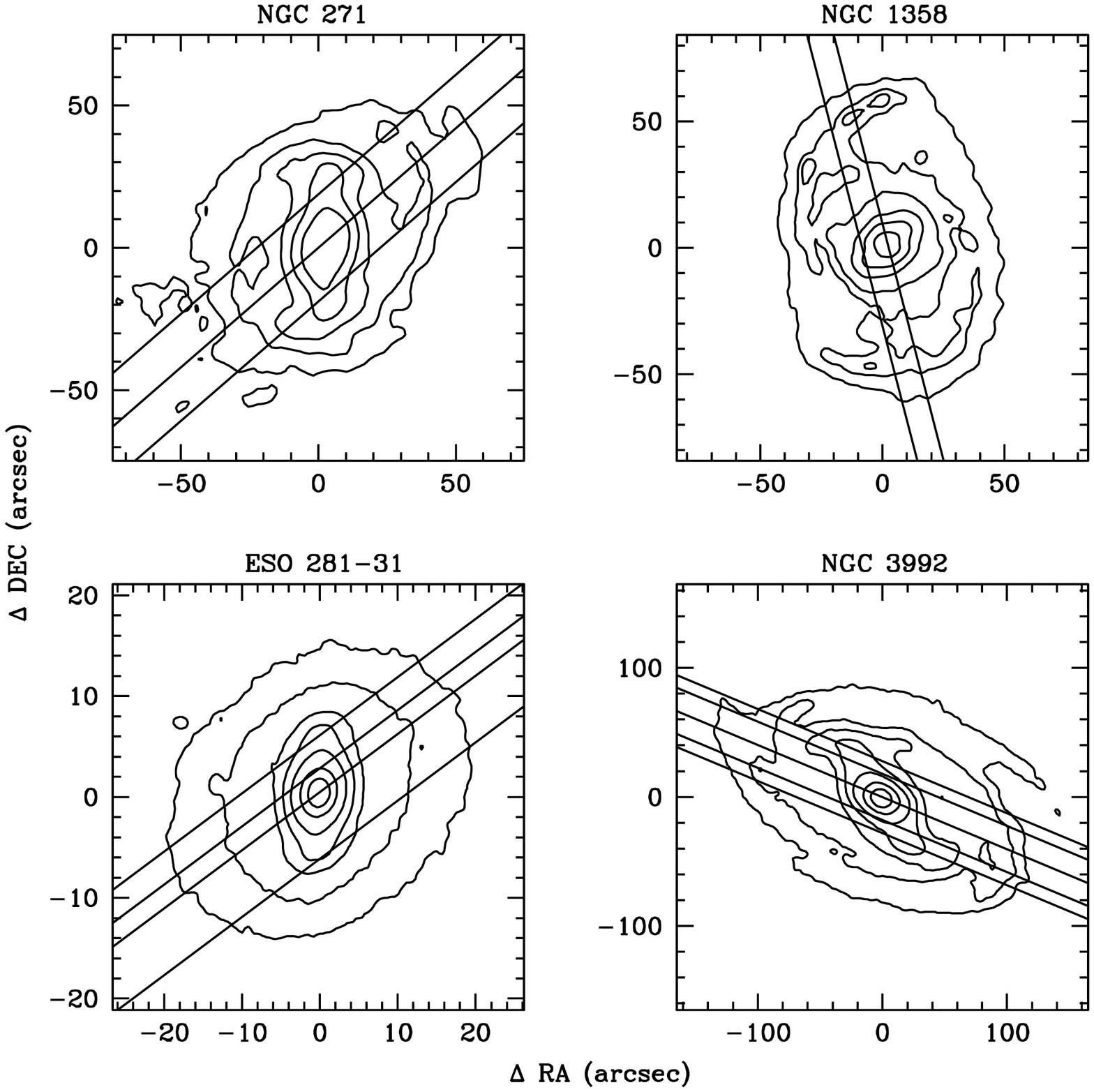}}
\ifsubmode
\vskip3.0truecm
\setcounter{figure}{0}
\addtocounter{figure}{1}
\centerline{Figure~\thefigure}
\else\figcaption{\figcapmosaic}\fi
\end{figure}


\clearpage
\begin{figure}
\epsfxsize=15.0truecm
\epsfysize=15.0truecm
\centerline{\epsfbox{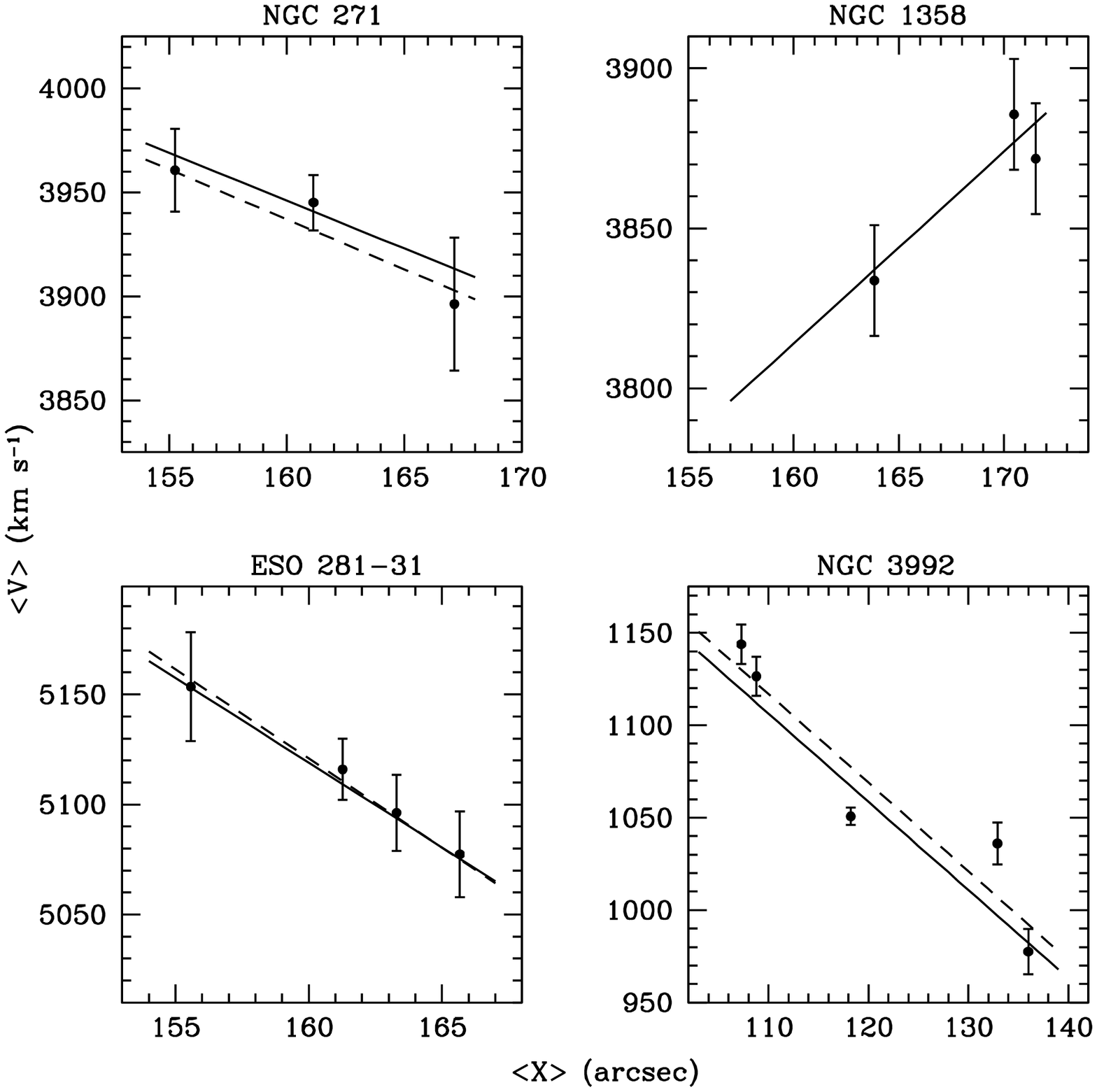}}
\ifsubmode
\vskip3.0truecm
\setcounter{figure}{0}
\addtocounter{figure}{1}
\centerline{Figure~\thefigure}
\else\figcaption{\figcapresults}\fi
\end{figure}


\clearpage
\begin{figure}
\epsfxsize=15.0truecm
\epsfysize=15.0truecm
\centerline{\epsfbox{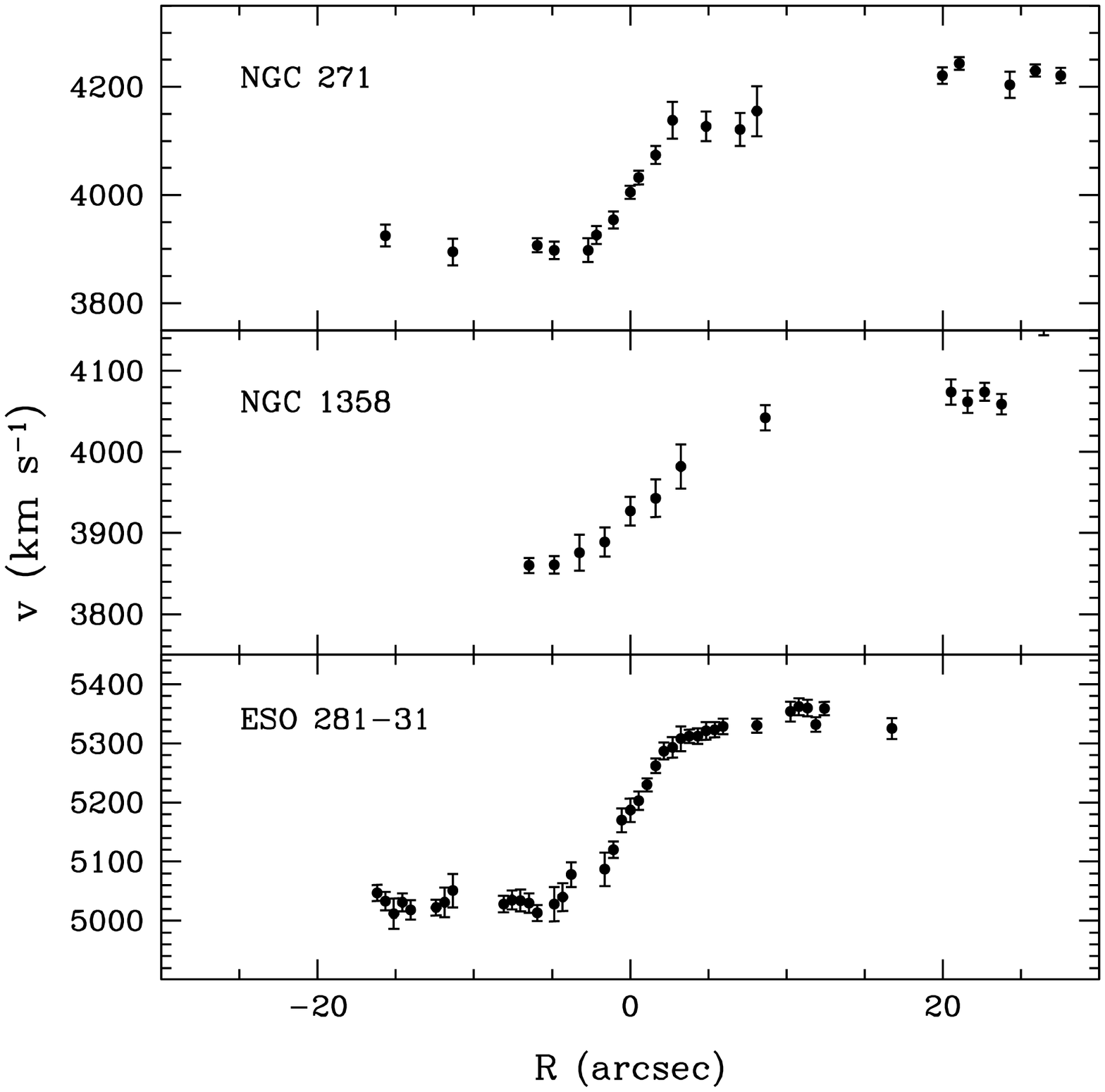}}
\ifsubmode
\vskip3.0truecm
\setcounter{figure}{0}
\addtocounter{figure}{1}
\centerline{Figure~\thefigure}
\else\figcaption{\figcapstellarkin}\fi
\end{figure}


\clearpage
\begin{figure}
\epsfxsize=15.0truecm
\epsfysize=15.0truecm
\centerline{\epsfbox{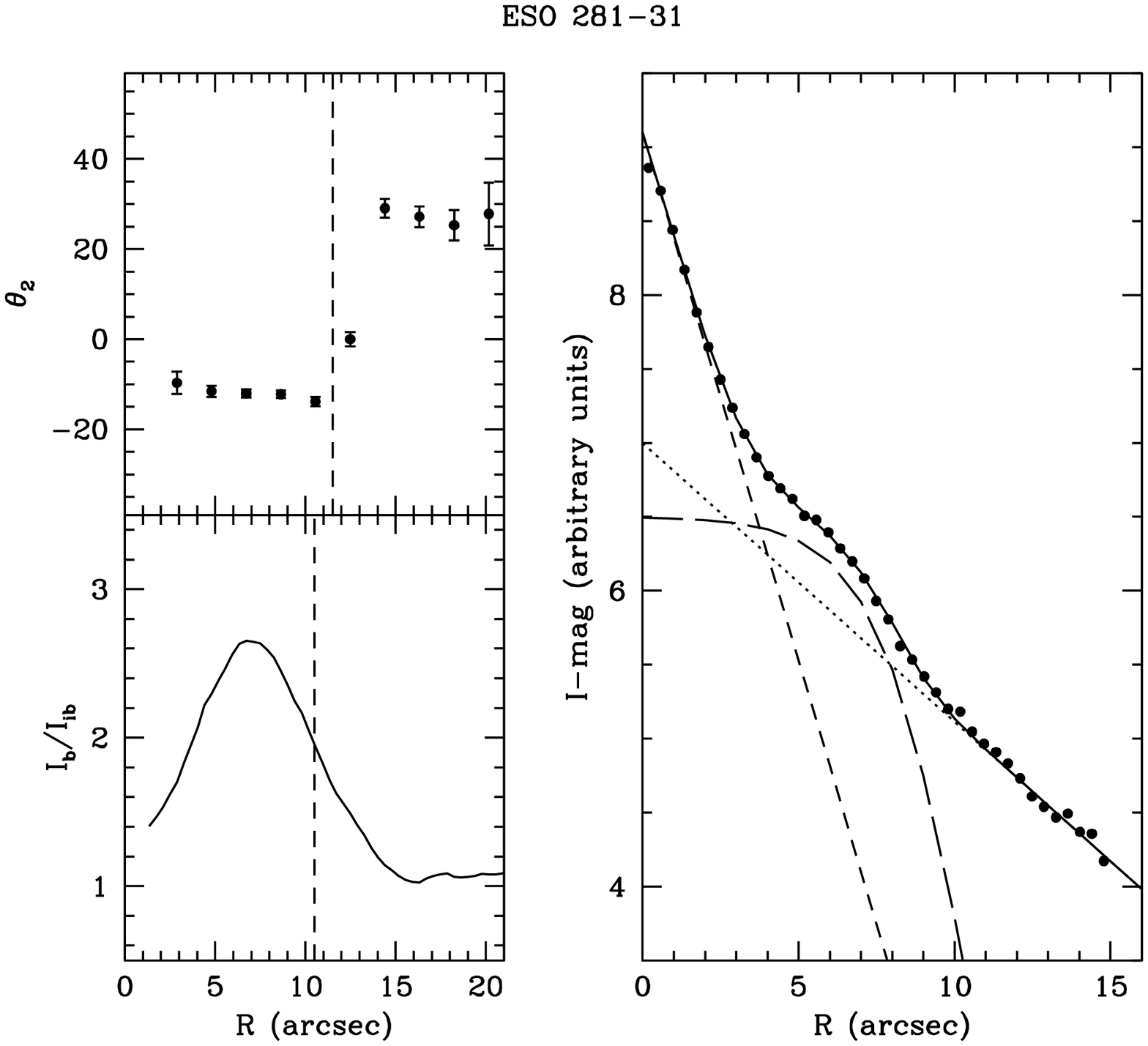}}
\ifsubmode
\vskip3.0truecm
\setcounter{figure}{0}
\addtocounter{figure}{1}
\centerline{Figure~\thefigure}
\else\figcaption{\figcapbarlength}\fi
\end{figure}


\clearpage
\begin{figure}
\epsfxsize=15.0truecm
\epsfysize=15.0truecm
\centerline{\epsfbox{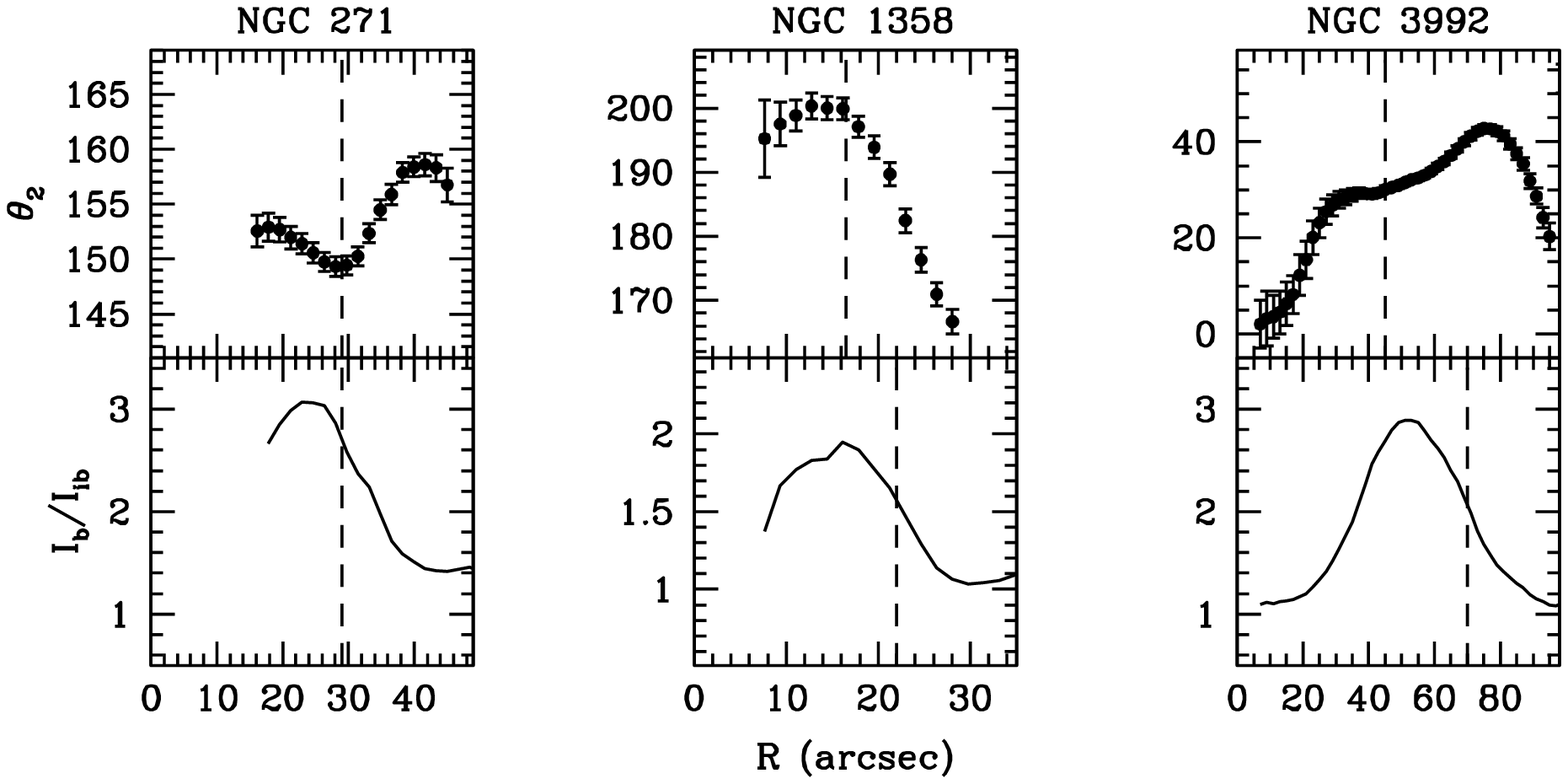}}
\ifsubmode
\vskip3.0truecm
\setcounter{figure}{0}
\addtocounter{figure}{1}
\centerline{Figure~\thefigure}
\else\figcaption{\figcapbarlengths}\fi
\end{figure}



\clearpage
\ifsubmode\pagestyle{empty}\fi

\begin{deluxetable}{lccccc}
\tablewidth{0pc}
\tablecaption{The barred galaxy sample}
\tablehead{
\colhead{Galaxy} & \colhead{Type} & \colhead{Redshift} & 
\colhead{PA$_{\rm disk}$} & \colhead{PA$_{\rm bar}$} & \colhead{Inclination}\\ 
\colhead{} & \colhead{} & \colhead{(km s$^{-1}$)} & \colhead{} & \colhead{}} 
\ifsubmode\renewcommand{\arraystretch}{0.68}\fi
\startdata
NGC 271    &    SBab  & 4129 &  130     &  178  & 36 $^{\rm a)}$ \\
NGC 1358   &    SB0a  & 4028 &  195     &  130  & 40 $^{\rm a)}$ \\
ESO 281-31 &    SB0   & 3459 &  110     &  173  & 47 $^{\rm b)}$ \\
NGC 3992   &    SBbc  & 1048 &  68      &   37  & 57 $^{\rm c)}$ \\
\enddata
\tablecomments{All velocities are from NED. \\
a) PAs are derived from DSS images, inclinations are from NED. \\
b) PAs and inclinations are derived from $I$ band photometry. \\
c) Data taken from Bottema \& Verheijen (2002).}
\label{t:sample}
\end{deluxetable}

\begin{deluxetable}{llc}
\tablewidth{0pc}
\tablecaption{Observing log} 
\tablehead{\colhead{Galaxy} & \colhead{Slit} & \colhead{$T_{\rm exp}$} \\
\colhead{} & \colhead{} & \colhead{(sec)}}
\ifsubmode\renewcommand{\arraystretch}{0.68}\fi
\startdata
NGC 271    & major axis	& 900   \\
           & NW	16''.0	& $2 \times 1800$  \\
	   & SE	16''.0	& $2 \times 1800$  \\
NGC 1358   & major axis & 900   \\
	   & NW 10''.0	& $2 \times 1800$  \\
ESO 281-31 & major axis & 900 + 1145  \\
	   & NW	3''.6	& $2 \times 1800$  \\
	   & NW 8''.6	& $2 \times 1800$  \\
	   & SE	8''.6	& 1800  \\		
NGC 3992   & major axis	& 900	\\
	   & NE	30''.0	& 1200 + 1800	\\
	   & NE 50''.0	& $2 \times 1800$  \\
	   & SW	30''.0	& 1200 + 1800	\\
	   & SW 50''.0	& $2 \times 1800$  \\	
\enddata
\label{t:obslog}
\end{deluxetable}

\begin{deluxetable}{lcccccc}
\tablewidth{0pc}
\tablecaption{Observationally derived parameters}
\tablehead{\colhead{Galaxy} &\colhead{$\Omega_p \sin i$} 
& \colhead{$\sigma_0$} & \colhead{$h$} & \colhead{$R_{\rm CO}$} & \colhead{$R_{\rm BAR}$} &
\colhead{$R_{\rm BAR-ALT}$} \\
\colhead{} & \colhead{($\kms {\rm arcsec}^{-1}$)}&\colhead{($\kms$)} &
\colhead{(arcsec)} & \colhead{(arcsec)} & \colhead{(arcsec)}
& \colhead{(arcsec)}}
\ifsubmode\renewcommand{\arraystretch}{0.68}\fi
\startdata
NGC 271    & 4.6 $\pm$ 2.5  & 200 $\pm$ 10 & 15.0 $\pm$ 1.0 & $44^{+30}_{-16}$  & 29 $\pm$ 1  & 28.8 \\
NGC 1358   & 6.0 $\pm$ 2.9  & 180 $\pm$ 10 & 4.5  $\pm$ 1.0 & $23^{+19}_{-7}$   & 19 $\pm$ 3  & 21.4 \\
ESO 281-31 & 7.7 $\pm$ 3.0  & 150 $\pm$ 10 & 5.3  $\pm$ 1.0 & $20^{+12}_{-4}$   & 11 $\pm$ 1  &  9.7 \\
NGC 3992   & 4.8 $\pm$ 0.3  &              &                & $45^{+3}_{-3}$    & 57 $\pm$ 12 & 67.8 \\
\enddata
\label{t:obspar}
\end{deluxetable}

\begin{deluxetable}{llcc}
\tablewidth{0pc}
\tablecaption{Results of the TW method}
\tablehead{\colhead{Galaxy} & \colhead{Type} & \colhead{$\Omega_p$} &
\colhead{$\mathcal{R}$} \\
\colhead{} & \colhead{} & \colhead{($\kms {\rm kpc}^{-1}$)} & \colhead{}}
\ifsubmode\renewcommand{\arraystretch}{0.68}\fi
\startdata
NGC 271    & SBab  & 25  $\pm$ 9   & $1.5^{+1.0}_{-0.5}$ \\
NGC 1358   & SB0a  & 31  $\pm$ 15  & $1.2^{+1.0}_{-0.4}$ \\
ESO 281-31 & SB0   & 27  $\pm$ 11  & $1.8^{+1.1}_{-0.4}$ \\
NGC 3992   & SBbc  & 73  $\pm$ 5   & $0.8^{+0.2}_{-0.2}$  \\
\enddata
\label{t:twresults}
\end{deluxetable}


\end{document}